%% file: meso2-rev.tex
\documentstyle[epsfig]{europhys}

\input euromacr

\begin{document}

\euro{}{}{}{}

\Date{}

\title{
Giant 
Tunneling Magnetoresistance, Glassiness, and the
Energy Landscape at Nanoscale Cluster Coexistence
}

\author{Sanjeev Kumar$^1$, Chandra Shekhar Mohapatra$^2$
 and Pinaki Majumdar$^3$}

\institute{
$^1$~Institute for Physics, Theoretical Physics III, Electronic
Correlations and Magnetism,\\
  University of Augsburg, 86135 Augsburg, Germany\\
$^2$~Department of Physics, I.I.T Kharagpur, Kharagpur 721 302, India\\
$^3$~Harish-Chandra  Research Institute,
 Chhatnag Road, Jhusi, Allahabad 211 019, India
}

\rec{}{}


\maketitle

\begin{abstract}

We present microscopic results on the giant tunneling magnetoresistance  
that arises from  the nanoscale coexistence of ferromagnetic metallic (FMM) 
and antiferromagnetic insulating (AFI) clusters in a disordered two dimensional
electron system with competing double exchange and superexchange interactions.
Our Monte Carlo study allows us to map out the different field regimes in  
magnetotransport and correlate it with the evolution of spatial structures.
At coexistence, the isotropic $O(3)$  model shows signs of slow relaxation, 
and has a high 
density of low energy metastable states, but no genuine glassiness.  However, 
in the presence of weak magnetic anisotropy, and below a field dependent 
irreversibility temperature $T_{irr}$, the response on field cooling (FC) 
differs distinctly from that on zero field cooling (ZFC). 
We map out the phase diagram of this `phase coexistence glass', highlight how 
its response differs from that of a standard spin glass, and compare our results 
with data on the manganites.

\end{abstract}

\section{Introduction}
First order phase transitions
involve a regime of metastability and 
phase coexistence. Two phases, either both ordered,
or one ordered and the other disordered, continue to be minima
of the free energy and 
the system can get trapped in the metastable minimum, or
be in a state of  macroscopic coexistence of the two phases. 
This scenario is complicated by the
presence of disorder. 
Disorder introduces preferential pinning or `nucleation'
centers for the two phases, and require us to consider the  energy 
$({\cal E})$
as a functional of the full order parameter
field, $\phi_{\bf r}$, say \cite{imry-ma,imry-wortis,dag-book}. 
The result is a 
pattern of coexisting clusters in real space, and 
a rugged 
landscape for ${\cal E}\{\phi_{\bf r}\} $
with many deep local minima in configuration space. 
The non trivial energy landscape results in slow
relaxation and history dependence in the response of
the system, the generic signatures of `glassiness'.
Such connection between phase coexistence and glassy effects  has
been explored in detail in vortex matter in superconductors 
\cite{glass-vortex1}, 
and more recently in the `colossal magnetoresistance' manganites 
\cite{mang-fc-zfc,mang-relax-prl,mang-pers-mem}.

Electron systems at coexistence, {\it e.g}, the manganites,
involve charge degrees of freedom, and, in addition
to glassiness, raise the possibility of 
a dramatic electrical response and insulator-metal
transitions.
It has been established that
there is coexistence of ferromagnetic metallic (FMM)
and charge ordered antiferromagnetic insulating (CO-AF-I) clusters
in the low $T_c$ manganites \cite{mang-clust-expt1,mang-clust-expt2},
notably the  La$_{1-x-y}$Pr$_y$Ca$_x$MnO$_3$
family. These materials  exhibit `out of equilibrium'
features like irreversibility \cite{mang-fc-zfc},
slow relaxation \cite{mang-relax-prl}, 
and persistent field memory \cite{mang-pers-mem}. 
In addition, since  conduction depends
on electron tunneling between ``half-metallic'' 
FMM clusters, a weak magnetic field greatly enhances the conductance 
\cite{mang-clust-expt1} by aligning the cluster moments. 
The field induced
suppression of resistivity $(\rho)$ can attain  $\rho(0)/\rho(h)
\sim 10^2$, at fields $h \sim 1$ Tesla,  down to zero temperature.
Coexistence, glassiness 
 and giant low temperature tunneling magnetoresistance (TMR)
 are intimately~related.

A first principles theory capturing nanoscale coexistence, TMR, and
the memory effects
requires a method  
which can handle multiple interactions, quenched disorder, and 
thermal effects simultaneously.
In this paper we use a recently developed Monte Carlo (MC) technique 
\cite{sk-pm-scr} to 
study a model of competing double exchange (DE) and superexchange (SE)
in the background of weak disorder, and clarify the following:
$(i)$~the origin and magnitude of
the large TMR at cluster coexistence and the 
high field evolution of  the magnetic state,  
and
$(ii)$~the nature of the  
``phase coexistence glass'', 
that arises  from a non trivial energy landscape and 
magnetic anisotropy, and its difference with respect to 
a canonical spin glass.

\section{ Earlier work} 
Models of competing DE and SE, in the presence of weak disorder,
have been studied in the recent past 
via `exact' simulation
on small lattices \cite{dag-clust1}.
These studies suggest spatial coexistence
of pinned clusters of the competing ordered phases. However,
being limited to very small sizes, 
they are unable to clarify
the cluster distribution, the transport properties, or possible
memory effects in the system.

In the absence of a microscopic approach, 
a resistor network phenomenology \cite{dag-clust1}
has been developed to model transport
at coexistence.
Although useful as a starting approximation, it is unable to
take into account
the `spin overlap' between clusters that controls magnetotransport, 
or the spatial correlations in the cluster distribution. 
We also know of no microscopic calculation  addressing the 
slow relaxation and glassiness in the coexistence regime.
In this paper we extend our earlier study  
at  zero field  \cite{sk-pm-meso1}
to clarify the magnetotransport and explore the 
history dependence 
in the coexistence regime.
\begin{figure}
\begin{center}
{
\epsfxsize=6.5cm \epsfysize= 6.5cm \epsfbox{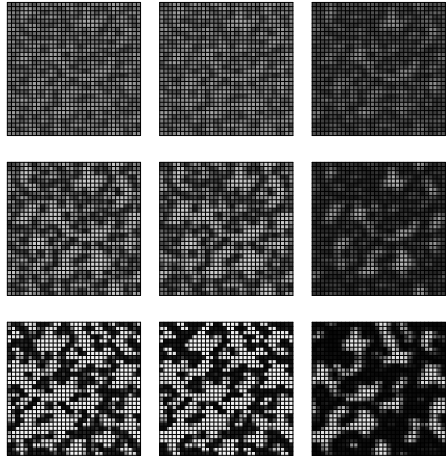}
}
\caption{ 
Cluster pattern for $n=0.1$ in the isotropic model. 
We  show the spatially resolved 
nearest neighbour spin correlation 
$f({\bf R}_i) = 
\langle {\bf S}({\bf R}_i).{\bf S}({{\bf R}_i  + {\hat x} }) \rangle$.
The system size is  $32 \times 32$.
Field variation along the row: $h = 0.0, 0.01, 0.10$, left to right,
and, down the column, the temperature are $T= 0.06, 0.03, 0.01$.
Dark regions are FMM, white are AF, grey are paramagnetic.}
\end{center}
\end{figure}

\section{Model and method}
We address the issues of TMR and `glassiness' using 
the following model \cite{sk-pm-meso1}
of competing DE and SE in
the presence of weak disorder in two dimension (2D):
\begin{equation}
H =
-t \sum_{\langle ij \rangle, \sigma}  
c^{\dagger}_{i \sigma} c^{~}_{j \sigma}
+  \sum_{i } (\epsilon_i - \mu) n_{i}  
- J_H\sum_i {\bf S}_i {\bf .} {\vec \sigma}_i 
+   J_S\sum_{\langle ij \rangle} {\bf S}_i.{\bf S}_j 
-h\sum_i  S_{iz} + H_{anis}
\end{equation}
The hopping is $t=1$ for nearest neighbours, and sets our basic 
energy scale,  and the structural disorder $\epsilon_i$ is
uniformly distributed between $\pm \Delta/2$. 
$J_H$ is the Hunds coupling,  
$J_S$ is the AF superexchange, 
and $h$ is the applied magnetic field.
We will use $H_{anis} = -\alpha\sum_i S_{iz}^2$, 
when studying the effect of anisotropy. 

We assume   
$J_H/t \rightarrow \infty$ and a classical core spin with
$\vert {\bf S}_i \vert =1$.
It is known that using finite $J_H/t$
\cite{dag-poib}, or a `quantum' $S=3/2$
core spin \cite{dag-poib} does not affect the principal
observations regarding FMM-AFI phase competetion.  
We also ignore the field coupling to electron spin since that
only renormalises the net moment.
Projecting  out fermion states at $+J_H/2$ we obtain
$ H_{el} 
= -t\sum_{\langle ij \rangle} f_{ij}
(~e^{i \Phi_{ij}}  \gamma^{\dagger}_i  \gamma_j + 
h.c~) + 
\sum_i (\epsilon_i - \mu) n_i $.
The $\gamma$ are spinless fermions, and the 
hopping amplitude, $g_{ij} = f_{ij} e^{i\Phi_{ij}}$, 
between locally aligned states,
can be written in terms of the polar angle $(\theta_i)$ and
azimuthal angle $(\phi_i)$ of the spin ${\bf S}_i$ 
as
$  cos{\theta_i \over 2} cos{\theta_j \over 2}$ 
$+
sin{\theta_i \over 2} sin{\theta_j \over 2}
e^{-i~(\phi_i - \phi_j)}$.
The magnitude  of the hopping is 
$f_{ij} = \sqrt{( 1 + {\bf S}_i.{\bf S}_j)/2 }$,
while the phase is specified by 
$tan{\Phi_{ij}} = Im(g_{ij})/Re(g_{ij})$.
We use the same technique as in  \cite{sk-pm-meso1} 
to construct an explicit classical spin Hamiltonian,  with self
consistently computed couplings, from the spin-fermion problem:
$ H_{eff} =
-\sum_{\langle ij \rangle} D_{ij} f_{ij}  
+ J_S\sum_{\langle ij \rangle} {\bf S}_i.{\bf S}_j
- h\sum_i  S_{iz}$.
The exchange $D_{ij}$ are  determined 
self consistently as the thermal average of    
${\hat \Gamma}_{ij} = 
(e^{i \Phi_{ij}}  \gamma^{\dagger}_i  \gamma_j + h.c)$
over configurations generated by $H_{eff}$.
The $D_{ij}$'s that emerge at consistency  
depend on temperature $(T)$ and  the magnetic field, 
and are  strongly 
inhomogeneous and spatially correlated. 
Our transport calculation  
is based on an exact implementation of the Kubo formula
described in detail elsewhere  \cite{sk-pm-transp}.

At weak  $J_S$, 
the competing phases in the clean problem are a FMM 
for $n \ge n_c$ (a critical density), 
and an AFI at $n=0$. There
is a discontinuous jump in $n$ from $n=0$ to $n=n_c$ 
across the AFI-FMM transition.
Disorder broadens the transition, leading to a regime of FMM-AFI cluster
coexistence. In 
this paper, as earlier \cite{sk-pm-meso1},
we set $J_S=0.05$ and 
$\Delta=1.0$,  choose
$n=0.1$ to be at 
the center of the coexistence regime, and focus  on the effects of 
finite $h$.
\begin{figure}
\begin{center}
{\epsfxsize=12.5cm \epsfysize=4.5cm \epsfbox{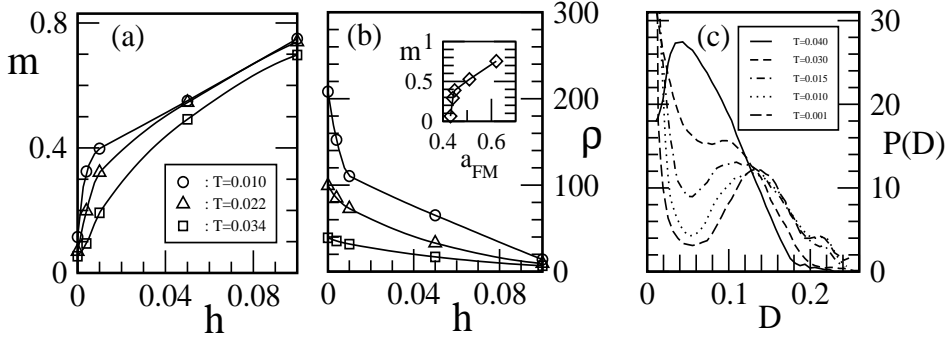}}
\caption{ 
Magnetism and transport in the isotropic model:  
$n=0.1$ and $\Delta = 1.0$.
$(a)$~$h$ dependence of magnetisation $m(h,T)$ 
after ZFC,  $(b)$~$h$ dependence of resistivity $\rho(h,T)$ 
after ZFC. 
Note the sharp low field response. The inset to $(b)$ shows the growth in
$m(h)$ correlated with the change in FMM `surface coverage'
at $T=0.01$.
Consistent with Fig.1, the  FMM 
regions hardly grow at low $h$, while $m(h)$ increases rapidly.
$(c)$~The bond distribution $P(D,T)$  obtained from
the self-consistent calculation. 
}
\end{center}
\end{figure}

\section{Results}
Our results are mainly in two parts: (A).~the effect of $h$ 
on transport and magnetisation in the fully self-consistent
isotropic model, and (B).~results in the presence of magnetic 
anisotropy, employing the spatially correlated bonds obtained
in (A).

(A).{\it~Fully self-consistent calculation without anisotropy:} 
Fig.1 shows the spatially resolved 
nearest neighbour spin correlation
$f_s({\bf R}_i) = 
\langle {\bf S}({\bf R}_i).{\bf S}({{\bf R}_i  + {\hat  x}}) \rangle$.
The dark regions have strong FM correlations, the white regions are AF,
while the grey regions are paramagnetic.
The pattern is shown  for three
combinations of $h$ and $T$. 
We interpret the spatial patterns as~follows.

$(i)$~In the high $T$  paramagnetic
phase (top row) $f_s$ is small on all links since $T$ is 
much larger than the exchange scales, and 
it  is only 
at $h=0.10$ (extreme right) that we see signs of magnetisation. 
The response  is $T$ or $h$  dominated, with the
magnetic correlations not playing any essential role.
$(ii)$~The ``contrast'' between FMM and AFI regions 
improves with cooling (central row) and the segregation 
into FMM and AFI clusters becomes apparent.
Low field, $h \sim 0.01$ (centre), does not have any perceptible effect on
$f_s$ compared to $h=0$ (left), but $h=0.10$ (right) leads to a large
magnetisation.  
$(iii)$~At the lowest $T$ (bottom row), 
the $h=0$ cluster pattern is very distinct with 
the electrons  confined to the FMM clusters. A low field
does not lead to a significant difference in the cluster pattern but
a {\it large change} in $m(h)$ as we discuss next.
Large field, $h=0.10$,
leads to a  visible 
increase in the FMM volume fraction.

The insensitivity of the cluster pattern to weak fields, at all $T$, should
be seen in conjunction with the response in magnetism, $m(h,T)$, and
transport, $\rho(h,T)$, in Fig.2.(a)-(b).
Within each FMM cluster the moments are aligned at low $T$, but the
overall $m(T)$ is small at $h=0$ due to the random orientation
of the ``cluster moments''. 
The  effect of a weak field, applied after cooling the system,
 is to orient the cluster moments
parallel to the field. Since the clusters typically have $>  10$ spins
coupled together even a weak field leads to a sharp increase in
the magnetisation.
In addition, the highly resistive state at $h=0$, which inhibited 
electron tunneling due to random orientation of the cluster moments,
now  rapidly turns conducting due to increasing  spin overlap
between clusters.
The rotation of cluster moments, and the resulting TMR, 
is the predominant low temperature {\it zero field cooled} (ZFC) effect
in this system.
\begin{figure}
\begin{center}
\epsfxsize=11.5cm \epsfysize=4.5cm \epsfbox{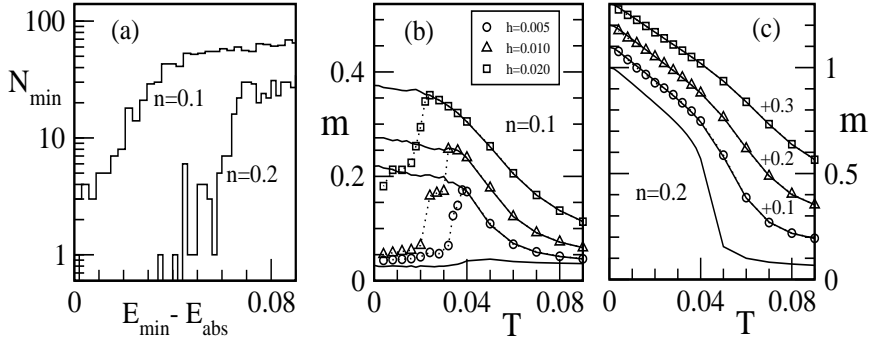}
\caption{ 
Metastable states, and memory effects in the anisotropic model:
$(a)$~The density, $N_{min}$, of metastable low energy minima
at $n=0.1$ (cluster coexistence) and
$n=0.2$ (almost homogeneous FMM) obtained in the isotropic model.
The energy $E_{min}-E_{abs}$ is measured with respect to the `absolute'
minimum  in the respective case. 
The results are qualitatively similar in the anisotropic model.
$(b)$~Magnetic response in the weakly anisotropic model, at $n=0.1$,
using bonds obtained from the self-consistent calculation. Firm
lines are for FC, dotted lines for ZFC.
$(c)$~FC-ZFC response in the 
$n=0.2$ (nearly homogeneous FMM) case, including anisotropy.
In this case the 
FC and ZFC response are identical, consistent with the
absence of low energy metastable states (panel $(a)$).
}
\end{center}
\end{figure}

When a larger field  is applied after cooling the system
two other effects come into play. The FMM phase grows in
volume, and residual spin disorder (due to finite $T$) 
{\it within the 
clusters} gets suppressed. The system now behaves almost 
like a canonical `clean'
DE ferromagnet,  
$m(h,T)$ nears
saturation and $\rho(h,T)$  drops to a low value controlled 
by the structural disorder.

Just as the spin correlations in Fig.1 evolve with changing $T$, the
bonds $D_{ij}$ also depend on temperature (through the self-consistency
condition). Fig.2.(c) shows the evolution of the 
system averaged bond distribution
$P(D)$ with changing $T$. Note that $P(D)$, although non trivial,
 in itself
has no information about the spatial correlation between bonds.

Using MC runs with equilibriation time 
$\tau_{eq} \sim 10^3$ steps revealed an apparent difference between
ZFC and FC in this system. However, with increasing $\tau_{eq}$ this
difference diminishes, and for $\tau_{eq} \ge 10^4$ the FC and 
ZFC response match.
It is known in the context 
of spin glasses that there cannot be any 
metastability in a $O(3)$ model in 2D \cite{sg-ref}. 
Nevertheless, the long
`relaxation time' (in the  MC sense) suggests that the effective
magnetic model probably has a non trivial energy landscape. 
To gain a deeper understanding of the observed relaxation we
examined the density of  low energy metastable 
configurations (LEMC), Fig.3.(a). We used the low $T$, $h=0$, energy 
functional ${\cal E} \{\theta, \phi\}  =
-\sum_{\langle ij \rangle} D_{ij} f_{ij}
+ J_S\sum_{\langle ij \rangle} {\bf S}_i.{\bf S}_j$, 
and employed  a conjugate gradient method to 
locate local minima starting from
random initial configurations.  Since this is an expensive effort for 
continuous spin systems, we restrict our results to  
 system size $8 \times 8$, 
using upto $10^4$ initial configurations.  
For the system at coexistence, $n=0.1$, the
density of LEMC is very high. We also studied the LEMC in our model
with electron density $n=0.2$ (an 
almost homogeneous FMM) 
and found that there is a clear gap between
the absolute minimum and the first  metastable state.
The high density of LEMC  at $n=0.1$ suggests that even weak
anisotropy might stabilise glassiness, 
which we discuss next.
\begin{figure}
\begin{center}
\epsfxsize=8.5cm \epsfysize=7.5cm \epsfbox{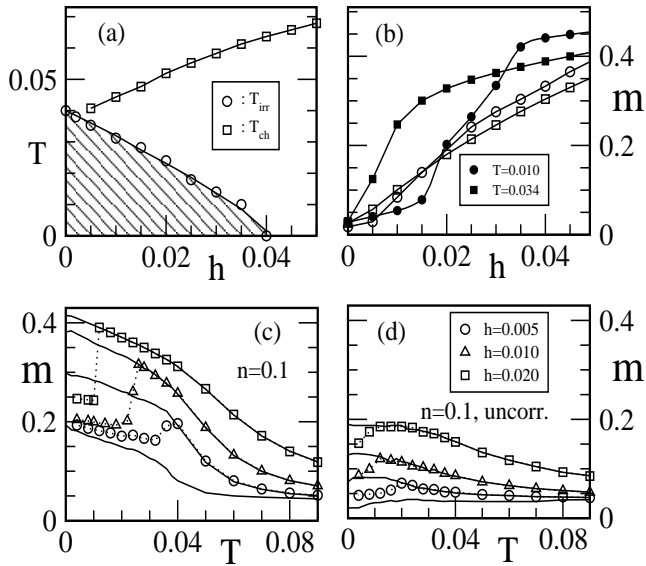}
\caption{ 
$(a)$~The $h-T$ phase diagram with anisotropy, $\alpha=0.02$.
The shaded area is the regime of irreversibility, bounded by
the irreversibility temperature $T_{irr}(h)$. The upper curve
refers to the `upturn' in the FM response.
$(b)$~Field response after ZFC 
computed with correlated bonds (filled symbols) and uncorrelated
bonds (open symbols).
$(c)-(d)$~Comparing FC-ZFC response between the PCG and homogeneous
SG,   
$(c)$~response obtained with 
correlated bonds for  the $n=0.1$ case, and
$(d)$~the model with {\it uncorrelated}  bonds $D^{SG}_{ij}$
picked from $P(D)$. All these systems have size
$24 \times 24$,  $J_S=0.05$, and 
uniaxial anisotropy with $\alpha=0.02$.
}
\end{center}
\end{figure}

(B).{\it~Anisotropic model with correlated bonds:}
The full self consistent calculation, involving iterative
bond evaluation, is difficult to perform in systems where
$\tau_{eq}$ is large. However, the qualitative character of
the  bond distribution,  Fig.2.(c), computed  self-consistently,
does not change for $T < 0.02$. Therefore,
to clarify the interplay of ``clustering'' 
({\it i.e}, spatial  correlation between the bonds) and
magnetic anisotropy we studied the following model via classical MC:
$$
H_{corr} = - \sum_{ij} D_{ij}^{corr} f_{ij} + 
J_S \sum_{ij} {\bf S}_i.{\bf S}_j
-h \sum_i S_{iz} + H_{anis}.
$$
The bonds $D^{corr}_{ij}$ are obtained from the 
self consistent scheme at low $T$. They are spatially correlated,
with the kind of spatial clustering seen in Fig.1.
This model ignores the thermal and field evolution of the $D_{ij}$
that is seen in the fully consistent calculation.

We studied the FC-ZFC difference in this model, Fig.3.(b), 
with bonds corresponding to $n=0.1$. 
Even with $\tau_{eq} > 10^4$ and averaging time $\tau_{av} > 10^4$,
the ZFC response differs clearly from the FC response.
Below a temperature $T_{irr}(h)$,
the FC and ZFC results are distinct,  with $m(T,h)$ being 
higher in the FC case.
The relative difference between  FC and ZFC  on $m(T,h)$,
at any $T$, decreases with increasing $h$, while the `irreversibility
temperature', $T_{irr}(h)$, 
reduces with increasing $h$, vanishing at $h \sim 0.04$.
For  the $n=0.2$ case 
 the FC and ZFC results are indistinguishable,
Fig.3.(c),
consistent with the simple energy~landscape.

Fig.4.(a) puts together the $h-T$ phase diagram that describes
our phase coexistence glass (PCG) in the presence of anisotropy,
with the shaded region denoting the regime of irreversibility.
The  difference between  FC and ZFC is 
well known in spin glasses \cite{sg-ref}, and 
has  also been observed  \cite{mang-fc-zfc} in the 
phase separated manganites. 
The key question, however, is  whether the glassiness
observed at coexistence can be understood in terms of 
the standard 
categories of $(a)$~a homogeneous spin glass (SG), or 
$(b)$~a `superparamagnet' (SP),
or $(c)$~a cluster glass (CG).
Let us analyse our spatial patterns and magnetic results
with reference to these models, as well as experimental data
on the manganites.
 
$(a)$~Since we do not compute time dependent responses to  characterise 
our glassy state 
we cannot compare directly with relaxational results known for spin
glasses, {\it etc}. 
We instead studied  a specific `spin glass' model, derived from
the coexistence problem: 
$
H_{SG} = - \sum_{ij} D_{ij}^{SG} f_{ij} + 
J_S \sum_{ij} {\bf S}_i.{\bf S}_j
-h \sum_i S_{iz} + H_{anis}.
$
In contrast to the PCG, the $D^{SG}_{ij}$  are picked 
{\it at random} from $P(D)$ without retaining any spatial 
correlations.
There are three features that distinguish this `SG' from the  PCG:
$(i)$~by construction there is no spatial clustering in this model
and patterns as in Fig.1 do not emerge, $(ii)$~the response in $m(h)$ 
after ZFC is much sharper in the PCG compared to the SG, 
Fig.4.(b), and $(iii)$~the overall magnitude of $m(T,h)$ and the
FC-ZFC difference are  significantly larger in the PCG compared to
the SG, Fig.4.(c)-(d).
These differences have their origin in the `clustering' 
that characterises the PCG so let us contrast 
our results with standard features of inhomogeneous systems,
a SP or a CG.

$(b)$~The usual superparamagnet is a collection of 
magnetic nanoparticles, with strong  
ferromagnetic coupling within each particle (or cluster),
while the coupling between clusters is 
negligible. 
The low $T$ weak field response of our  PCG (in terms of
domain rotation) has some similarity with a SP, but 
at intermediate
field or temperature, see Fig.1, the correlation
 between `clusters' is
important.  

$(c)$~A cluster glass extends the SP picture to incorporate 
intercluster interaction. 
The detailed physical effects in a CG 
depend on the density of clusters and
their size distribution \cite{cg-ref} but  
the basic identity of the clusters,
and the coupling between them, do not depend on $T$ or $h$.
The  phenomenology of a cluster glass,
in terms of spatial correlation
and sharp low field response, 
is closest to our results. 
The differences are:
$(i)$~the entities in the PCG are elementary spins, albeit `clustered',
but not predefined composite objects, and $(ii)$~ the  ``cluster 
distribution''  in the PCG, as well as 
the coupling between
clusters, depends on temperature and field, effectively tuning the DE-SE
competetion. That does not happen in a CG.

Our model captures some of the key aspects of field response and glassiness
at coexistence in the manganites: the field regimes
observed in Pr$_{0.7}$Ca$_{0.3}$MnO$_3$ \cite{mang-fc-zfc} including
low field anisotropy, followed by domain rotation and then growth 
of the FM regions, are all seen in our Fig.2-3. We can also
capture the behaviour of $T_{irr}(h)$. 
We do not have data on the time dependence of magnetic relaxation,
as measured in
\cite{mang-relax-prl}, but our results provide insight on how the DE-SE
interaction evolves in a clustered system with variation in temperature. 
Finally, in contrast to bulk measurements, which can only hypothesise
about spatial structures, we can correlate 
the bulk properties with the underlying microscopic situation.

To conclude,
we have studied a model of competing double exchange and
superexchange in a weakly disordered 2D electron system and characterised
the glassy features of the resulting magnetic state. Our results describe 
how magnetic anisotropy and 
the non trivial energy landscape  at phase coexistence
combine to generate several of the glassy features observed in the 
manganites.
It would be interesting to study the 
detailed time dependent response 
of this model in 3D to clarify 
the outstanding memory 
effects \cite{mang-relax-prl,mang-pers-mem}
observed in the manganites.

\vspace{.2cm}

We acknowledge use of the Beowulf cluster at H.R.I.
SK  gratefully acknowledges support by the Deutsche
Forschungsgemeinschaft through SFB484.

{}


\end{document}

%% file: euromacr.tex
\newif\ifboo \boofalse

\def\Review#1{\boofalse{\it #1},}

\def\Vol#1{\ifboo Vol. {\bf #1}\else{\bf #1}\fi}
\def\Year#1{\ifboo #1\else(#1)\fi}

\def\Page#1{\ifboo {\rm p. #1}\else{\rm #1}\fi}